\documentstyle[aps]{revtex}

\begin{document}

 


\title{An unconventional Fermi liquid model for the optimally doped \\
and overdoped cuprate superconductors}

\bigskip
\author{George Kastrinakis}
\address{Department of Physics, University of Illinois at Urbana-Champaign,
1110 West Green St.,Urbana, IL 61801 }
\date{March 15, 1997}

\maketitle
\begin{abstract}
Based on an unconventional Fermi liquid model, we present several results 
on the optimally doped and overdoped cuprate superconductors.
For the normal state, 
we provide an analytic demonstration, backed by
self-consistent Baym-Kadanoff (BK) numerical calculations, of the linear in $T$
resistivity and linear in 1/$\epsilon$ optical conductivity,
provided the interacting Fermi liquid has strong peaks in its density
of states (van-Hove singularities in 2 dimensions) near the chemical potential
$\mu$.
Moreover, we find that the interactions tend to pin these strong density
of states peaks close to $\mu$. 
We show that the {\em low} energy dependence of $\chi_{MMP}$ has a
fermionic origin.
We obtain particularly high transition temperatures $T_c$ from our 
BK-Eliashberg scheme
by introducing an {\em ansatz} for the fermionic susceptibility of the carriers.
We postulate that the latter is {\em enhanced} in an additive 
manner due to the weak antiferromagnetic order of the CuO$_2$ planes.
We have obtained a $d_{x^2-y^2}$ gap with $T_c > 120 ^o$K for
n.n. hopping $t=250meV$.
\end{abstract}

\vspace{.4cm}


	The nature of the many-body state of the cuprate superconductors 
is a core question for the understanding of these materials 
\cite{Scala,agl,pines}.
In this Letter, we show that a {\em minimum} unconventional Fermi liquid
model accounts in a natural, comprehensive and internally consistent manner
for several normal state characteristics. The introduction of an {\em
ansatz} for the susceptibility of the carriers further allows us to obtain 
particularly high transition temperatures $T_c$.

	Our starting point is the 2-dimensional Hamiltonian
\begin{equation}
H = \sum_{k,\sigma} \epsilon_k \; c_{k,\sigma}^{\dag} c_{k,\sigma}
+ \frac{A_o}{2} \sum_{k,k',q,\sigma,\sigma'} c_{k+q,\sigma}^{\dag} 
c_{k'-q,\sigma'}^{\dag} c_{k',\sigma'} c_{k,\sigma} 
+ \frac{U_o}{2} \sum_{k,k',q,\sigma} c_{k+q,\sigma}^{\dag} 
c_{k'-q,-\sigma}^{\dag} c_{k',-\sigma} c_{k,\sigma}  \;\;.
\end{equation}
$c_{k,\sigma}^{\dag}$ is an electron creation operator and 
$\epsilon_k = -2 t (\cos{k_x}+\cos{k_y})-4t'\cos{k_x}\cos{k_y}-
2t''(\cos{2k_x}+\cos{2k_y})$.
$A_o$ is the strength of a contact Coulomb interaction, and $U_o$ the strength
of a paramagnon interaction.
A small {\em negative} $A_o$ would mimic an isotropic phononic interaction.

We have developed a self-consistent many-body treatment of the above
Hamiltonian. We write a set of diagrams within the frame of a
Baym-Kadanoff
conserving approximation \cite{bk}, summing bubble and ladder diagrams 
\cite{gk2}.
There is a free energy
functional $\Phi[G]$ of the Green's function $G$, such that the self-energy
$\Sigma$ is given by the relation $\Sigma = \delta \Phi[G]/\delta G$.
We thus obtain a set of
self-consistent equations for $G(k,\epsilon_n)$ and $\Sigma(k,\epsilon_n)$ 
:
\begin{equation}
G(k,\epsilon_n) = \frac{1}{i \epsilon_n + \mu - \epsilon_k - \Sigma(k,\epsilon_n)}  \;\;, 
\;\;
\Sigma(k,\epsilon_n) = - \frac{T}{N^2}\sum_{q,\omega_m} V(q,\omega_m) 
\; G(k-q,\epsilon_n-\omega_m) \;\; .
\end{equation}
$\mu$ is the chemical potential and the Matsubara frequencies 
are $\epsilon_n=(2n+1)\pi T$ and $\omega_m=2m\pi T$ for fermions and bosons, 
respectively. 
This system of equations is solved numerically, similarly to e.g.
\cite{flex,mont,Grabo,vilk,carb,Andersen,dahm}.
We work with a given number $M$ of Matsubara frequencies
and a $N\times N$ discretization of the Brillouin zone
($M=256-480$ and $N \geq 64$).
The potential
$V(q,\omega_m)=V_{ex}(q,\omega_m)-V_H(q,\omega_m) \;\;,
V_{ex}(q,\omega_m) = \frac{A_o - (A_o+U_o)^2 \chi_o(q,\omega_m)}
{1 - (A_o+U_o)^2 \chi_o^2(q,\omega_m)} \;\; 
\frac{1}{1-A_o^2 \chi_o^2(q,\omega_m)} \; - \; A_o  \;\;, 
V_H(q,\omega_m) = \frac{(A_o+U_o)^3 \chi_o^2(q,\omega_m)}
{1 - (A_o+U_o) \chi_o(q,\omega_m)} + \frac{A_o^3 \chi_o^2(q,\omega_m)}
{1-A_o \chi_o(q,\omega_m)}  \;\;
$ \cite{gk2}.
$V_{ex}(q,\omega_m)$ includes 
vertex corrections due to the internal dressing of bubbles with $A_o$. 
Our potential is an extension of the fluctuation-exchange 
(FLEX) approximation of Bickers, Scalapino and White \cite{flex},
as can easily be seen by taking $A_o \rightarrow 0$.
Moreover, $\chi_o(q,\omega_m)
= - (T/N^2) \sum_{k,\epsilon_n} G(k+q,\epsilon_n+\omega_m) G(k,\epsilon_n) \; .
$

	All the convolution operations are done by using the Fast Fourier
Transform (FFT), in order to cut down calculation time. 
	We use Pad\'e approximants \cite{pade} to analytically continue 
our results to the real frequency axis.

	The numerical solution of the many-body system yields always
an interesting result for the self energy at finite temperature. 
$Im \Sigma(k,\epsilon)$ turns out 
to be essentially linear in energy in an interval 
$\epsilon_1<\epsilon<\epsilon_2$. 
However it does have the correct parabolic Fermi liquid bevahior 
for $\epsilon \rightarrow 0$.
Furthermore,
the energy interval of linear behavior expands as the energy $\epsilon_{vH}$
at which are located the
van-Hove singularities (vHs) at the points 
$q_o =(\pm \pi,0), (0,\pm \pi)$
approaches the 
chemical potential $\mu$.

	We give an analytic derivation of this result. At finite temperature 
$Im \Sigma(k,\epsilon)$ is given by \cite{agd} :
\begin{equation}
Im \Sigma^R(k,\epsilon) = \sum_{q,\omega} Im G^R(q,\epsilon-\omega) 
Im V^R(k-q,\omega) \; 
\{\coth(\omega/2T) \; + \; \tanh((\epsilon-\omega)/2T) \}\;\;.
\end{equation}
Taking 
$Im G^R(k,\epsilon) = -\pi \delta(s_{k,\epsilon}), \; \; 
s_{k,\epsilon} = \epsilon+\mu-\epsilon_k \;\; ,
$
we obtain
\begin{equation}
Im \Sigma^R(k,\epsilon) = -\pi \sum_q Im V^R(k-q,s_{q,\epsilon}) \;
\{ \coth(s_{q,\epsilon}/2T) \;
+ \; \tanh((\epsilon_q-\mu)/2T) \}\;\; .
\end{equation}
Setting $Im G(k,\epsilon)$ equal to a delta function is a reasonable
approximation for this purpose, since numerically $Im \Sigma(k,\epsilon)$ is 
{\em very small} compared to the band energy for small couplings. 
Then, the difference of $Im \Sigma(k,\epsilon)$, as seen in our numerical
calculation, for small and large coupling constants is {\em quantitative}
rather than qualitative.

	We take
$Im V^R(q,x) = \sum_{n=0}^{\infty} \; V_q^{(2n+1)}(0) \; x^{2n+1}/(2n+1)!$.
This is true for an electronically mediated interaction,
with a polarization which is a regular function of $\omega$ (see also below).

	Now, it is easy to show analytically \cite{gk} that in the 
low $T$ limit for $\mu > \epsilon_{vH}$
\begin{equation}
Im \Sigma^R(k,\epsilon) = -\pi a_k (\epsilon + c) \;\;, 
\epsilon_1 <  \epsilon  < \epsilon_2 \;\;,
\end{equation}
where $a_k = \sum_{q \sim q_o} V_{k-q}^{(1)}(0)$, $c = \mu - \epsilon_{vH}$,
and $\epsilon_1 = \mu - \epsilon_{vH}$, 
$\epsilon_2 = \epsilon_c + \epsilon_{vH} - \mu $, and 
$\epsilon_c \leq 8t$ is a characteristic energy.


	Similarly \cite{gk}, in the high temperature limit 
$T > (\mu-\epsilon_{vH})/4$ \cite{varel} we obtain 
\begin{equation}
Im \Sigma^R(k,\epsilon) = - 2 \pi a_k T \;\;.
\end{equation}
The result is the same for $\mu<\epsilon_{vH}$, save for a multiplication by 2
of the r.h.s. of eq. (5).
We note that the $T$ and $\epsilon$ dependence of the result are {\em independent}
of $k$ - {\bf thus leading necessarily to a linear in $T$ resistivity and a 
linear
in $1/\epsilon$ optical conductivity,} even with inclusion of vertex corrections
in the calculation.

	The overall behavior of $Im V(q,\omega)=Im \chi_o(q,\omega)
\; |\varepsilon(q,\omega)|^2$, where $\varepsilon(q,\omega)$ is a properly
defined dielectric function, follows from any screened 
interaction between the carriers. Hence the argument for the
linear in energy and temperature behavior of the scattering rate 
$\tau^{-1}(T,\epsilon)$
is equally generic. It relies essentially on a large coefficient for the 
linear in energy term of $Im \chi_o$, and the presence of vHs
{\em near} the Fermi surface. What is more, in our numerical solution 
we observe that the energy $\epsilon_{vH}$ of the singularities is pushed 
by the interactions close to $\mu$. The shape of the 
imaginary part of the self-energy $Im \Sigma(k,\epsilon)$ of the interacting 
system
being responsible for the modification of the density of states $N(\epsilon)$, 
through the relation $N(\epsilon)=-Tr \; Im G(k,\epsilon)/\pi$.
Namely, $Im \Sigma(k,\epsilon)$ has a peak below $\mu$ and a dip above it, which 
account for the transfer of the spectral weight.
This feedback effect reinforces the r\^{o}le of the singularities, which 
are more effective in producing a linear scattering rate the closer they 
are to the Fermi surface. This seems to be a plausible explanation
for the common characteristic of a good many cuprates whose vHs
are located between 10-30 $meV$ {\em below} the Fermi surface
\cite{bednorz} (see also \cite{bsk}).

	It is interesting that the electron doped 
Nd$_{2-x}$Ce$_x$CuO$_{4+\delta}$
whose vHs are at 
approximately $\mu$-350 meV \cite{stanford},
has a usual Fermi liquid $\tau^{-1}(T) = const. \; T^2$ \cite{tsuei}.
This lends support to the picture described above.

	Markiewicz has given a review of related work in the frame of 
the so-called van-Hove scenario \cite{mark}. 



	The celebrated Millis-Monien-Pines susceptibility\cite{pines}
\begin{equation}
\chi_{MMP}(q,\omega) = \frac{X_1 \; \xi^2}{1+\xi^2 (q-Q)^2 - 
i \omega/\omega_{SF}} \;\;,
\end{equation}
has been used to fit the {\em low} energy part of the 
susceptibility of the 
cuprates in both NMR rate and inelastic neutron scattering (INS) experiments.
The short range antiferromagnetic order, a remnant of the parent 
antiferromagnetic materials, with correlation length $\xi$, 
is responsible for the peak of the susceptibility for $q$ near $Q$.
Typically $\xi$ is of the order of the lattice constant ($\xi$ decreases
as the doping increases), while
$\omega_{SF} \approx 10-40 meV$.

	The origin of the small magnitude of $\omega_{SF}$ has remained 
elusive thus far.
E.g. Chubukov, Sachdev and Sokol have interpreted it as a damped spin
wave mode \cite{css}. Spin waves are clearly well-defined excitations
in underdoped cuprates.
However to date there is no experimental proof that they do so for small
energies in the normal phase of the optimally doped and overdoped regimes. 

	An alternative explanation is a fermionic origin for $\omega_{SF}$.
First, let us consider the {\em total} susceptibility put forward
by Onufrieva and Rossat-Mignod\cite{onu}
\begin{equation}
\chi_t(q,\omega) = \frac{\chi_{AF}(q,\omega) + \chi_F(q,\omega)}
{1 - J_q (\chi_{AF}(q,\omega) + \chi_F(q,\omega))} \;\;.
\end{equation}
$\chi_{AF}(q,\omega)$ is due to the localized Cu spins, which interact via 
the effective spin exchange $J_q$, and $\chi_F$ is a purely
fermionic susceptibility.
This $\chi_t(q,\omega)$ encompasses in an appealing way the idea of the
entangled carrier-spin dynamics in the cuprates.
	Now, we use $\chi_t$ above with
\begin{equation}
\chi_{AF}(q,\omega) = \frac{\chi_1 \; \xi^2 }{1+\xi^2 (q-Q)^2 - f(\omega)} 
\;\;, \;\;
\chi_F(q,\omega) = \chi_o (1+i \omega/\omega_o + O(\omega^2)) \;\;,\;\;
q \rightarrow Q  \;\;.
\end{equation}
If $|f(\omega)|\ll \omega/\omega_o$ and $1 \gtrsim J_Q \xi^2 \chi_1$
we recover essentially $\chi_{MMP}(q,\omega)$ - which is
itself an {\em approximate} form of the true susceptibility - with 
$\omega_{SF} \rightarrow \bar\omega \propto \omega_o$.

	The origin of small $\omega_o(\vec q)=\vec \nabla \epsilon_{k_F} \vec 
q$ 
\cite{pino} is the proximity of the vHs to the Fermi surface 
- c.f. the discussion in the
context of the scattering rate above. From the numerical solution
of our system, we easily obtain values of $\omega_o$ comparable to 
the experimentally relevant ones, when the chemical potential is
near $\epsilon_{vH}$. 
Moreover, when $\epsilon_{vH}$ approaches $\mu$, 
$\omega_o$ is quickly suppressed towards zero.
As already mentioned, the small difference $\mu-\epsilon_{vH}$ is found
in several cuprates - see \cite{bednorz}.
It has also been suggested \cite{bsk} 
that this characteristic may be true irrespective of the doping,
as long as the latter is appropriate for superconductivity.
It is then clear that the
carriers residing in the vicinity of the vHs, with 
a large density of states, give rise to a {\em small} $\omega_o$ 
as discussed above. Hence it is very 
interesting to know 
how universal this band-structure characteristic of the cuprates is,
as it may explain naturally the magnitude of $\omega_o$.
This idea can be tested experimentally by e.g. determining by
INS the value of $\bar\omega \propto \omega_o$ for 
Nd$_{2-x}$Ce$_x$CuO$_{4+\delta}$, which is expected to be enhanced
as a result of the big difference between $\epsilon_{vH}$ and $\mu$.

	Moreover, we thereby propose that the {\em effective} non-interacting
susceptibility of the carriers is given by the following {\bf ansatz}
\begin{equation}
\chi_o(q,\omega) \rightarrow \chi_o^{eff}(q,\omega) = \chi_o(q,\omega) + 
a \; \chi_{AF}(q,\omega) \;\;.
\end{equation}
$\chi_o$ and $\chi_{AF}$ are as above and 
$0 < a < 1$ is a weighting factor 
(which should in general depend on doping, band-structure etc.). 
The intuitive idea is that the carriers
should feel the ordered antiferromagnetic background of the CuO$_2$ planes, 
even in the absence
of phase separation \cite{ek,cn}.
In this way, the fermionic susceptibility acquires an antiferromagnetic
enhancement, which may then influence pairing, through an electronically
mediated interaction - see below. 
Note that with this effective $\chi_o^{eff}$ our many-body 
approximation remains {\em conserving}, since the relation $\Sigma = \delta F/
\delta G$ is valid, as $\chi_{AF}$ does {\em not} depend on $G$.

Further,
$\chi_o^{eff}$ can explain the temperature dependence of the Hall resistivity
of the cuprates \cite{stp}, 
as it can produce an effective carrier potential $V$ peaked at $Q$.

	To locate $T_c$ we solve
the 
gap equation
\begin{equation}
\Delta(k,\epsilon_n)=-\frac{T}{N^2}\sum_{k',\epsilon_n'} V_{p}(k-k',\epsilon_n-\epsilon_n') 
|G(k',\epsilon_n')|^2 \Delta(k',\epsilon_n')	\;\;.
\end{equation}
The pairing potential $V_{p}
= V_{x}(q,\omega_m)  + V_h(q,\omega_m)  \;\;,  
V_{x}(q,\omega_m) = \frac{A_o+U_o}{1 - (A_o+U_o)^2 \chi_o^2(q,\omega_m)} \;\;
\frac{1}{1+A_o \chi_o(q,\omega_m)} \;\;,  
V_h = \frac{(A_o+U_o)^2 \chi_o(q,\omega_m)}
{1 - (A_o+U_o) \chi_o(q,\omega_m)} \;\;
$ \cite{gk2}.

The highest $T_c$'s correspond to a $d_{x^2-y^2}$ gap, and are obtained 
by including an
antiferromagnetically enhanced susceptibility in the calculation, following our
{\em ansatz} eq. (10).
We consider two similar forms for the AF susceptibility, namely (A) :
$\chi_{AF}^A(q,\omega_m) = X_o \; \sum_{i=1}^4 \Gamma_i^{-1}\; 
\theta(\omega_c - |\omega_m|)$, where $\Gamma_i=\xi^{-2} + (q-Q_i)^2$,
and (B) : 
$\chi_{AF}^B(q,\omega_m) = X_o \; D \; \sum_{i=1}^4 \{ \omega_m - (2/\pi) 
\omega_m \arctan(\omega_m/D) - \Gamma_i D + (2/\pi) 
\Gamma_i D \arctan(\Gamma_i) \} / \{\omega_m^2 - (\Gamma_i D)^2\}$, 
with $Q_i=(\pm\pi,\pm\pi)$ and $D$ being a cut-off frequency, 
above which $Im \chi_{AF}(q,\omega)=0$.
Form (B) has appeared in \cite{millis}. Here the characteristic spin wave
frequency $\omega_S \propto \xi^{-z}$, and the $z=2$ scaling regime
has been assumed.

With $t=250 meV$, $A_o=0$ and $\xi=1$ (increasing the latter leads to a 
reduction of $T_c$) we have the following results.
For form (A) we obtain $T_c \simeq 125^o K$, for $a X_o=0.2eV^{-1}$,
$\omega_c=4t$, $t'=-0.11t$, $t''=0.5t$, $U_o=1.27eV$ and $n=0.88$.
For form (B) we obtain $T_c \simeq 105^o K$, for $a X_o=0.5eV^{-1}$,
$D=16t$, $t'=-0.11t$, $t''=0.25t$, $U_o=1.47eV$ and $n=0.91$.

We note that the basic effect of considering moderate values $A_o<0$ 
is to allow for greater 
values of $U_o$ before the AF instability condition $U_{tot} \chi_o = 1$,
with $U_{tot} = A_o + U_o$, sets-in
(recall that we take $\chi_o^{eff}$ peaked at $Q_i=(\pm\pi,\pm\pi)$). 
Hence a finite phonon interaction 
allows for a bigger on-site/paramagnon coupling.

There is an {\em optimum} value of $a$ for the maximum attainable $T_c$. 
This is due to the form of the pairing potential $V_p$ above : 
the AF instability condition allows $U_{tot} \chi_o^{eff} < 1$ only.
Now, for a given value of the latter product, the highest $V_p$ - which 
in principle yields the highest $T_c$ as well - 
will correspond
to the highest possible $U_{tot}$. This in turn corresponds to a smaller 
$\chi_o$, and hence to a {\em small} but finite optimum $a$.
In fact, Zheng et al. \cite{zheng} noticed that the highest $T_c$ 's are 
obtained
for a {\em combination} of both optimum total carrier concentration
and of a reduced relaxation rate $(1/T_1)_{Cu} \propto
T \; \lim_{\omega \rightarrow 0} \sum_{q} A_q^2 \; 
Im \chi_{AF}(q,\omega)/\omega $,
in properly normalized units (here $A_q$ is the hyperfine coupling).
These conclusions are in accordance with our picture, which yields both
special values of the filling factor $n$, as well as special {\em small} 
values of the product
$a \chi_{AF}$ as a prerequisite for the highest attainable $T_c$ 's.

Monthoux and Pines have obtained similarly high transition temperatures
\cite{mont} with their approach. 

	We note that the vHs at $q_o$ suppress a 
$s$-wave gap if the pairing potential $V_p>$0 ($V_p$ defined above 
is negative for sufficiently large and negative $A_o$ and/or $U_o$, 
and positive otherwise).
The $s$-wave gap of Nd$_{2-x}$Ce$_x$CuO$_{4+\delta}$
can be due either to $V_p<$0 for some relevant parts of the phase space
in this material, or to the fact that the vHs are 350$meV$
below the Fermi surface, and hence ineffective here, or possibly to both facts.
Yet another possibility which may coexist with the above characteristics
is that $V_p$ is strongly peaked close to zero momentum owing to 
the band structure of this material. All these factors can
result in the $s$-wave $T_c$ being higher than the $d$-wave $T_c$, and hence in 
the dominance of the former over the latter.

	In brief, let us discuss the possible connection to the physics
of the underdoped cuprates, which are believed to be in a phase separated
regime, with AF domains of spins separated by stripes of holes \cite{ek,cn}.
One can envisage that with doping increasing towards the optimal regime,
the stripes melt into an effective Fermi liquid, and the physics described
here is recovered.

	To summarize, we present a single plane Fermi liquid model
which for the normal state can explain the salient transport properties, 
the low energy
dependence of $\chi_{MMP}$, and their relation to the existence
of van-Hove singularities close to the Fermi surface. We also obtain 
enhanced transition temperatures $T_c$. Special attention is paid
to Nd$_{2-x}$Ce$_x$CuO$_{4+\delta}$, the properties of which,
despite appearances, 
we believe to be fully consistent with those of the majority of cuprates.

	The author has enjoyed discussions with 
Yia-Chung Chang, Gordon Baym, Joseph Betouras,
Girsh Blumberg, Nejat Bulut, Antonio Castro Neto, Lance Cooper, Tony Leggett, 
Dmitrii Maslov, Philippe Monthoux, 
J\"{o}rg Schmalian, Raivo Stern, Qimiao Si and
Branko Stojkovic.
This work was supported by the Research Board of
the University of Illinois, the Office of Naval Research under
N00014-90-J-1267 and NSF under DMR-89-20538.


\end{document}